\documentclass[preprint,12pt,3p]{elsarticle}

\usepackage{amssymb}

\usepackage{graphics}
\usepackage{graphicx}
\usepackage{comment}
\usepackage{caption}
\usepackage{subfig}
\usepackage{float}
\usepackage{hyperref}

\usepackage{setspace}
\usepackage{subfloat}

\usepackage{multirow}

\usepackage{lineno}

\journal{Nuclear Instruments and Methods}

\begin{document}

\newcommand{\cf}{$^{252}$Cf~}
\newcommand{\Can}{$^{13}$C($\alpha$,n)$^{16}$O~}
\newcommand{\Lipn}{$^{7}$Li(p,n)$^{7}$Be~}

\begin{frontmatter}

\title{Light Response of Poly(ethylene 2,6-napthalate) to Neutrons }

\author[label1,label2,labelc]{Brennan Hackett\corref{cor1}}
\author[label3]{Richard deBoer}
\author[label1,label2]{Yuri Efremenko}
\author[label2]{Michael Febbraro}
\author[label2]{Jason Nattress}
\author[label3]{Dan Bardayan}
\author[label3]{Chevelle Boomershine}
\author[label6]{Kristyn Brandenburg}
\author[label3,label5]{Stefania Dede}
\author[label6]{Joseph Derkin}
\author[label3]{Ruoyu Fang}
\author[label6,label7]{Adam Fritsch}
\author[label3]{August Gula}
\author[label4]{Gy\"urky Gy\"orgy}
\author[label6]{Gulakhshan Hamad}
\author[label6]{Yenuel Jones-Alberty}
\author[label3]{Beka Kelmar}
\author[label3]{Khachatur Manukyan}
\author[label3]{Miriam Matney}
\author[label3]{John McDonaugh}
\author[label3]{Shane Moylan}
\author[label3]{Patrick O'Malley}
\author[label3]{Shahina Shahina}
\author[label6]{Nisha Singh}

\address[label1]{Department of Physics and Astronomy, University of Tennessee, Knoxville Tennessee, USA}
\address[label2]{Oak Ridge National Laboratory, Oak Ridge, TN, USA}
\address[labelc]{Max Planck Institute for Physics, Garching, Germany}
\address[label3]{Department of Physics and the Joint Institute for Nuclear Astrophysics, Notre Dame, Indiana 46556, USA}
\address[label6]{Edwards Accelerator Laboratory, Department of Physics and Astronomy, Ohio University, Athens, Ohio 45701, USA}
\address[label5]{Cyclotron Institute, Texas A\&M University, College Station, Texas, United States, 77843}
\address[label7]{Department of Physics, Gonzaga University, Spokane, Washington 99258, USA}
\address[label4]{Institute of Nuclear Research (ATOMKI), Hungarian Acadamy of Sciences, Debrecen, Hungary}

\cortext[cor1]{Corresponding author}

\begin{abstract}
There is an increasing necessity for low background active materials as ton-scale, rare-event and cryogenic detectors are developed.
Poly(ethylene-2,6-naphthalate) (PEN) has been considered for these applications because of its robust structural characteristics, and its scintillation light in the blue wavelength region.
Radioluminescent properties of PEN have been measured to aid in the evaluation of this material.
In this article we present a measurement of PEN's quenching factor using three different neutron sources; neutrons emitted from spontaneous fission in \cf, neutrons generated from a DD generator, and neutrons emitted from the \Can and the \Lipn nuclear reactions.
The fission source used time-of-flight to determine the neutron energy, and the neutron energy from the nuclear reactions was defined using thin targets and reaction kinematics.
The Birks' factor and scintillation efficiency were found to be $kB = 0.12 \pm 0.01$~mm MeV$^{-1}$ and $S = 1.31\pm0.09$ ~MeV$_{ee}$~MeV$^{-1}$ from a simultaneous analysis of the data obtained from the three different sources.
With these parameters, it is possible to evaluate PEN as a viable material for large-scale, low background physics experiments. 

\end{abstract}


\end{frontmatter}


\doublespacing

\section{Introduction}
\label{intro}

The next generation of rare-event physics experiments continue to pursue quasi-background free environments in ton-scale detectors~\cite{ProtoDUNE_2019, DarkSide-20k:2017zyg}.
This is being achieved through improvements in spatial and temporal background rejection, particle identification capabilities, higher active veto system efficiency, and developments in ultra-low background materials.

\begin{figure}
    \centering
    \includegraphics[width=0.5\textwidth]{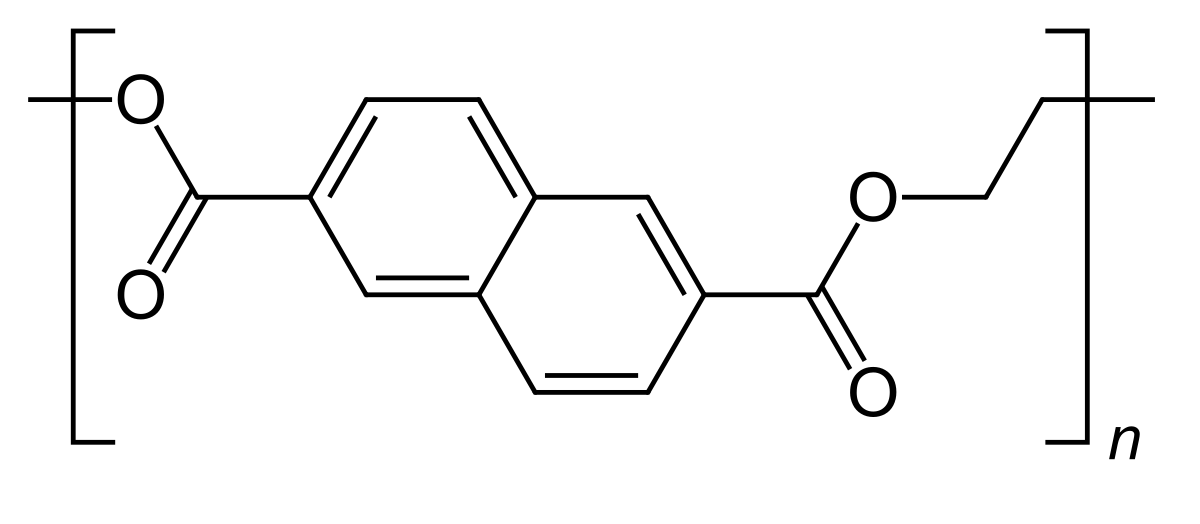}
    \caption{Skeleton structure of the repeating unit for the polymer PEN.}
    \label{fig:PENskel}
\end{figure}

Poly(ethylene 2,6-naphthalate) (PEN), is of interest as a low background, active material in rare-event and multi-ton liquid argon experiments~\cite{Manzanillas:2022zyh}.
PEN is an aromatic polyester (Fig. \ref{fig:PENskel}) that exhibits excellent mechanical strength and chemical resistance~\cite{Boulay:2021njr}. 
Additionally this material scintillates and wavelength shifts vacuum ultraviolet (VUV) light~\cite{Abraham:2021otn}. 
PEN is a unique polymer in that it scintillates inherently, with its primary scintillation light in the blue region~($\sim\!\!400$~nm), therefore, unlike traditional plastic scintillators, no additional fluors are required~\cite{PEN_selfveto}. 

For active veto applications in low background experiments, it is important to fully understand PEN's ability to veto radioactive decays, which may occur inside or on the surface of a PEN component (e.g. products of radon decays).
This can be understood through Monte Carlo simulations of the radioactive decays and the light propagating in the active veto system, allowing for the full characterization of the active veto system performance.
The light response for PEN has been measured for alpha particles~\cite{nakamura2014detection}, but never for neutron interactions. 
Quenching in scintillators is defined by considering the light produced by a nuclear recoil, $L(E)_{nr}$ as a fraction of the light produced by an electronic recoil of equivalent energy, $L(E)_{er}$:

\begin{equation}
\label{qf}
    QF(E) = \frac{L(E)_{nr}}{L(E)_{er}}.
\end{equation}

There are many models to describe how the light is quenched as a function of energy in a scintillator~\cite{QF_plastic_scint}. This paper will focus on the use of Birks' law~\cite{BirksJ.B.JohnBetteley1967Ttap}: 

\begin{equation}
\label{birks}
    \frac{dL}{dr} = S \frac{\frac{dE}{dr}}{1+kB\frac{dE}{dr}}.
\end{equation}
Birks' law describes how the light per unit length, $\frac{dL}{dr}$, is quenched as a function of stopping power, $\frac{dE}{dr}$. 
The scintillation efficiency $S$, and the Birks' constant, $kB$, describes the shape of the light response curve, and to what degree it is non-linear as a function of energy deposition by the ionising particle. 
By numerical integration, it is possible to fit Eq.~\ref{birks} to $L(E)_{nr}$ to derive $S$ and $kB$ for proton recoils as a result of neutron interactions.  

To determine the quenching factors and determine the Birks' parameters, it is necessary to determine which measured events are from neutrons scattering and generating a nuclear recoil, and which are electron recoils from $\gamma$-ray interactions. 
It has been demonstrated that it is possible to separate these events in PEN with the use of pulse shape discrimination (PSD) as it allows the distinction between particles with different charge/mass ratios, such as electron and proton recoils ~\cite{PEN_selfveto}.
Eq.~\ref{eq:PSD} defines the PSD parameter: 
\begin{equation}
    \label{eq:PSD}
    PSD = \frac{\int_{tail} L.dt}{\int_{total} L.dt}
\end{equation}
Where $L$ is light, $dt$ is representative of the time of the pulse, $tail$ describes the decaying tail of the light pulse, and $total$ refers to the integral of the whole pulse. 
Optimal PSD separation was observed with a total integral of 430 ns, and an offset of 70 ns for the tail integral. 
This fits well with the previously measured time constant of PEN of 25 ns~\cite{Manzanillas:2022zyh}.

This paper is organized as follows; Section \ref{materials} provides details about the PEN scintillation detector used for this measurement; followed by Section \ref{experiment} which describes the experimental preparation and procedure for the neutron measurements. 
It also includes a description of the calibration method for the measurements, this is followed by a brief description of the error analysis method in Section \ref{analysisandresults}. 
Finally, Section \ref{conclusion} presents the findings for the quenching factor and Birks' constant for PEN and summarizes the results of this study.

\section{PEN scintillation detector} 
\label{materials}

\begin{figure}[h!]
    \centering
    \includegraphics[width=0.5\linewidth]{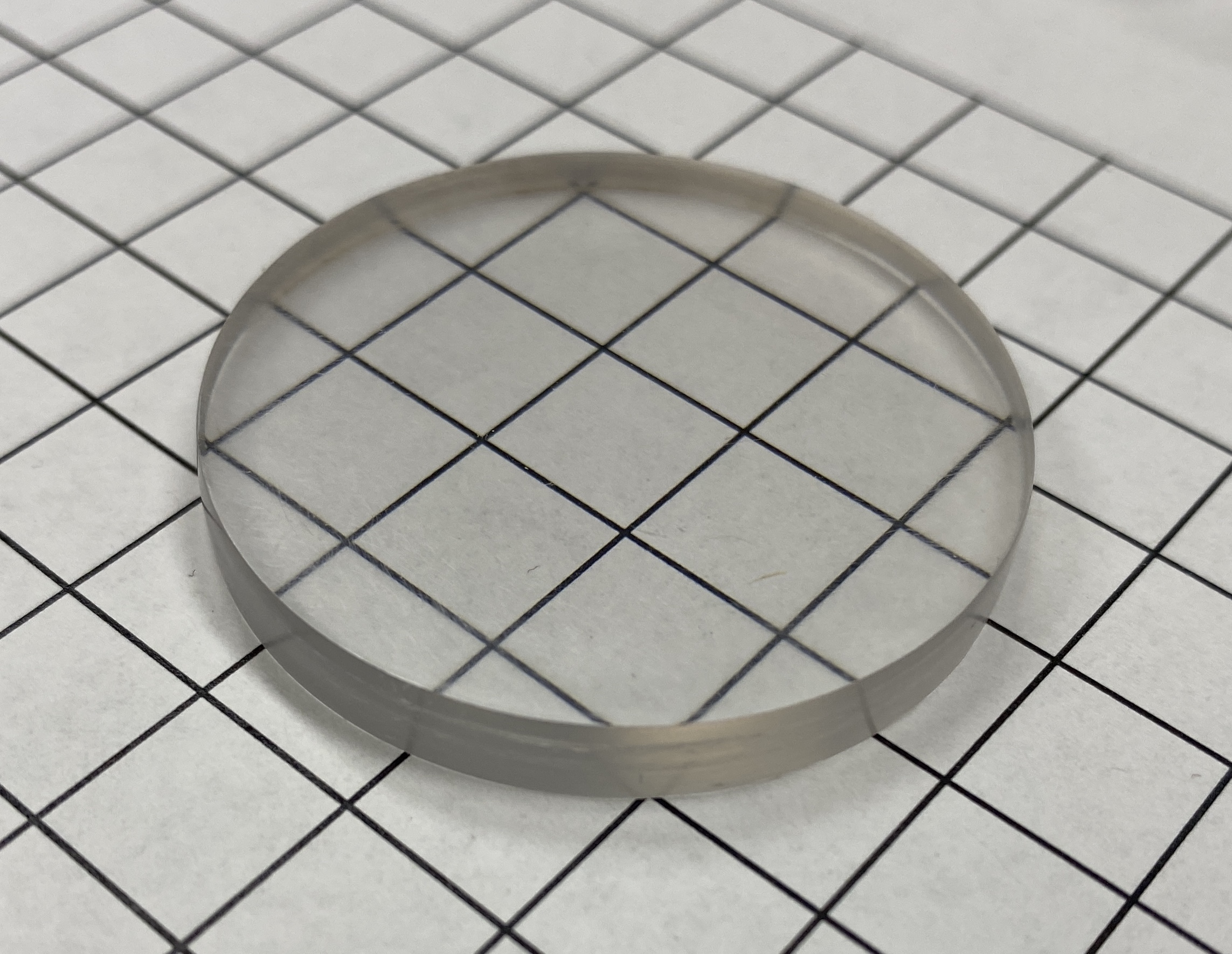}
    \caption{The PEN component used for these luminescent measurements has a radius of 5.08 cm and a thickness of 0.5 cm. }
    \label{fig:PEN_piece}
\end{figure}

PEN is a semi-crystalline aromatic polymer which co-exists in both the crystalline and amorphous phases~\cite{PEN_selfveto}.
The PEN component used for this measurement is an amorphous sample produced by an injection molding process described in Ref.~\cite{PEN_selfveto}. 
The PEN material used to produce this component was sourced from Teonex, batch TN-8065S, with a diameter of 5.08 cm and a thickness of 0.5 cm~(Fig.~\ref{fig:PEN_piece}). 

The detector consisted of three main components; a 3D printed cap to hold the PEN component and optically seal the detector, the PEN component itself, and a photomultiplier tube (PMT). 
For the measurement conducted using nuclear reaction neutrons, a Hamamatsu R6233 2" PMT was used. 
This tube is a hybrid design with box-and-grid technology to maximize light collection, and linear focused technology to improve pulse linearity~\cite{HamPMT}. 
For the measurement conducted using spontaneous fission neutrons, a Hamamatsu R7224 2" PMT was used. 
This tube has a linear focused design, with a smaller transient time spread and larger gain, resulting in a lower light threshold and improved time resolution, and therefore energy resolution. 
The same detector was used for both the spontaneous fission measurement and the DD generator measurement. 
Both detectors were lined with Teflon which serves as the optical reflector, and the component was optically coupled to the face of the PMT using Saint Gobain optical coupling compound, BC-630.

\section{Experimental Method} 
\label{experiment}

The experimental method was as follows; $L(E)_{er}$ was calibrated for each detector with gamma-ray sources and fixing the quenching factor for electronic recoils to 1. 
Next, the energy of the incident neutrons were determined using reaction kinematics or time-of-flight.
Finally, the quenching factor was determined by measuring the light produced for a proton recoil of a given neutron energy, $L(E)_{nr}$.

\subsection{Detector Calibration}

Gamma-ray radiation sources were used to calibrate the PEN scintillation detector in each measurement. 
These sources included $^{133}$Ba (356 keV), $^{22}$Na (511 keV), ${137}$Cs (661 keV) and $^{60}$Co (1121 keV and 1332 keV). 
The detector energy resolution and calibration constant was then determined by reproducing the experimental spectrum shape from simulated data. 
This was first achieved by fitting the edge of the gamma source data with a Gaussian function, defined in Eq.~\ref{gauss}. 
\begin{equation}
    \label{gauss}
    f(x) = k*\exp^{-\frac{1}{2}\left(\frac{x-\mu}{\sigma}\right)^2}
\end{equation}
The simulated data was then reshaped using a resolution and smearing function, defined in Eq.~\ref{FWHM} and Eq.~\ref{errorfunc}. 
\begin{equation}
    \label{FWHM}
    FWHM = E*\sqrt{\alpha^2+\frac{\beta^2}{E}}
\end{equation}

\begin{equation}
    \label{errorfunc}
    E' = (E + {erf}^{-1}(r)*\sqrt{2\left(\frac{FWHM}{2.35}\right)^2})*C.
\end{equation}
The parameters $\alpha$ and $\beta$ are constants that are particular for any specific scintillator and photo-multiplier combination and $C$ refers to a constant that is the inverse of the calibration factor. 
An iterative method was  applied to find optimal parameters for $\alpha$, $\beta$ and $C$ to reproduce the spectral shape defined by the initial Gaussian fit ($\mu$ and $\sigma$) of the experimental data edge.
A minimum $\chi^2$ is found between the fit function and smeared simulated deposited energy spectrum, and used to quantify the quality of the fit.

\begin{figure}
    \centering
    \includegraphics[width=0.7\linewidth]{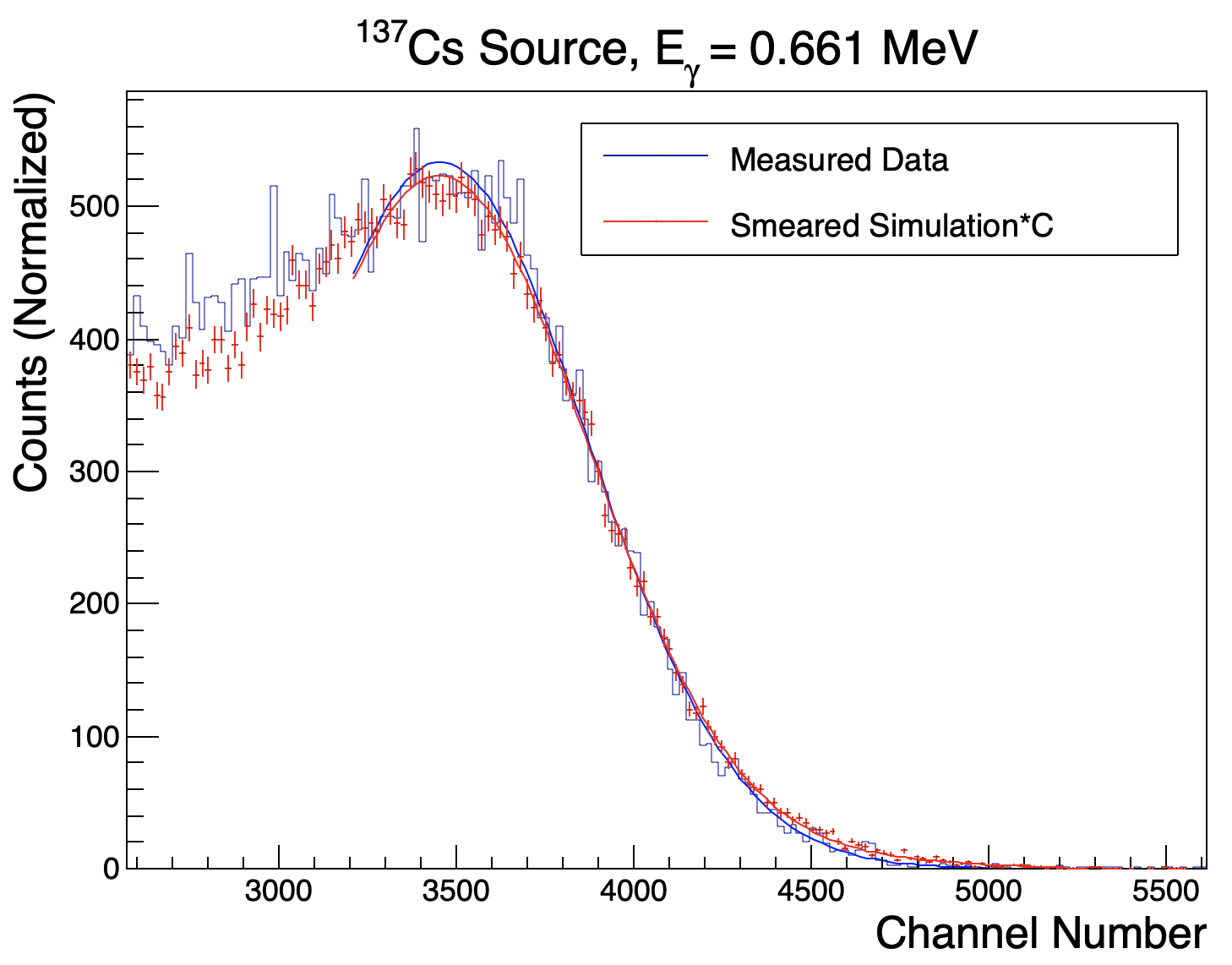}
    \caption{ The measured data of a $^{137}$Cs source with $E_{\gamma} = 0.661$~MeV, impinging on a PEN scintillation detector is shown in blue, and fit with a Gaussian (Eq.~\ref{gauss}). The simulation is shown in red, also fit with a Gaussian, and the statistics are normalized to the data. The Gaussian smearing (Eq.~\ref{FWHM} and Eq.~\ref{errorfunc}) is applied with the following values; $\alpha=0.053$, $\beta=0.12$ and $C=8020$.}
    \label{fig:gammadata}
\end{figure}

\begin{figure}
    \centering
    \includegraphics[width=0.7\linewidth]{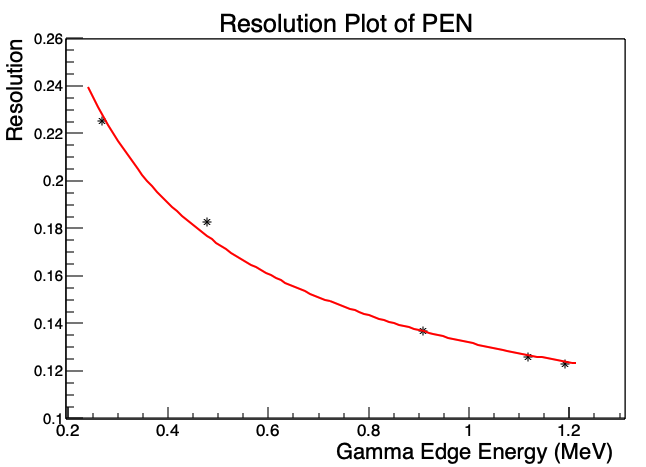}
    \caption{The detector energy resolution was derived by fitting various gamma-ray source spectra. This plot shows the detector resolution for the detector used at the University of Notre Dame for the nuclear reaction neutron sources.}
    \label{fig:gammares}
\end{figure}

An example is shown in Fig.~\ref{fig:gammadata}, where the simulated shape is compared with experimental data after smearing with the following values; $\alpha=0.053$, $\beta=0.12$ and $C=8020$.
The detector energy resolution is plotted as a function of energy in Fig.~\ref{fig:gammares}.

\subsection{Calibration of Neutron Sources and Experimental Setup}

Any single neutron source measurement could result in systematic errors in the measured light response, and would be limited in its energy range. 
However, by measuring the luminescent response of PEN using multiple neutron sources, as well as using different light collection systems and data acquisition systems, agreement between measurements will reduce the likelihood of global systematic errors. 
Therefore three different types of neutron sources were used; a white \cf neutron source with time-of-flight, monoenergetic neutrons from a DD generator, and monoenergetic neutrons produced from the \Can and \Lipn reactions using monoenergetic particle beams, made incident on thin targets where the beam experienced very small energy losses.
There are different methods used to determine the neutron energies between these three measurements.
The first is time of flight, where the kinetic energy of the neutron $E_n$ is determined by the flight time, $T$, the distance it travels, $D$, and the mass of the neutron, $m_n$. 
The neutron's energy and energy resolution of the measurement is defined by Eq.\ref{eq:nkin} and Eq.\ref{kineticenergy}.
\begin{equation}
    \label{eq:nkin}
    E_n = \frac{1}{2}m_n \left(\frac{D}{T} \right) ^2
\end{equation}
\begin{equation}
\label{kineticenergy}
    \frac{\Delta E_n}{E_n} = 2\sqrt{\left({\frac{\Delta D}{D}}\right)^2 +\left(\frac{\Delta T}{T}\right)^2}.
\end{equation}
The energy resolution of this measurement is defined by the uncertainty in the time and distance. 
It is possible to increase the flight path to improve the energy resolution, but this comes at the cost of statistics.
Additionally, as the neutron energy increases, its time-of-flight decreases, therefore resulting in a decrease in energy resolution. 
This places an upper limit on the energies that can be reasonably measured using time-of-flight.

The second method used to determine the neutron's energy is to use reaction kinematics. 
Neutrons produced in two-body nuclear reactions will have a well defined energy depending on the beam bombarding energy and the angle of emission, with an energy uncertainty determined by the solid angle acceptance, energy loss through the target and fluctuation in the beam energy. 
Both the beam energy fluctuations and the energy loss through the target were determined to be negligible.   
There are other reaction channels that produce large numbers of background events. 
PSD was used to remove background events by distinguishing between proton and electron recoils. 
The separation becomes worse at lower energies and therefore has limited ability to distinguish nuclear and electronic recoils at energies ${E_n<0.7}$~MeV.

\subsubsection{\cf Spontaneous Fission Source} \label{cf_section}

For this measurement, \cf was used as a white neutron source. The non-relativistic kinetic energy of the neutrons was determined using time-of-flight from the source to the detector. 
Neutrons are produced by spontaneous fission in \cf.
When a fission spontaneously occurs, both neutrons and photons are released. 
The \cf source used in this experiment is inside a small fission chamber, which will trigger on fissile material being released, and therefore will set a start time for the event~\cite{ORNL_FC_manual}. 
The \cf source is plated on a platinum disk and is set parallel to the signal collector. 
The \cf source and signal collector are set in a hemispherical ionization chamber filled with a mixture of $97\%$ Ar and $3\%$ CO$_2$ gas.

\begin{figure}
    \centering
    \includegraphics[width=0.7\textwidth]{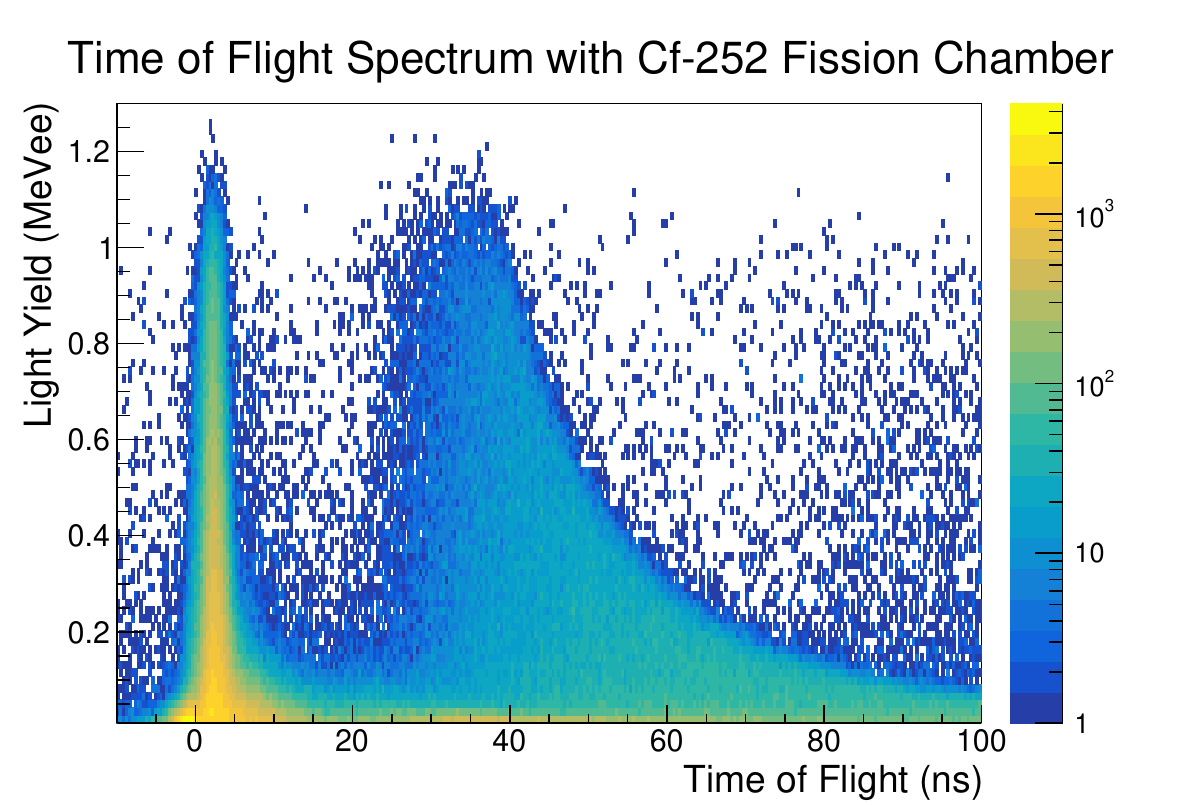}
    \caption{The time of flight was used to determine the energy of neutrons from the \cf source. This plot shows the time of flight or $\delta T$ for events from the fission chamber to the PEN scintillation detector versus the light yield for that event. The gamma flash can be seen focused around $\sim3$ ns. }
    \label{fig:TOF_PLOTS}
\end{figure}

\begin{figure}
    \centering
    \includegraphics[width=0.7\textwidth]{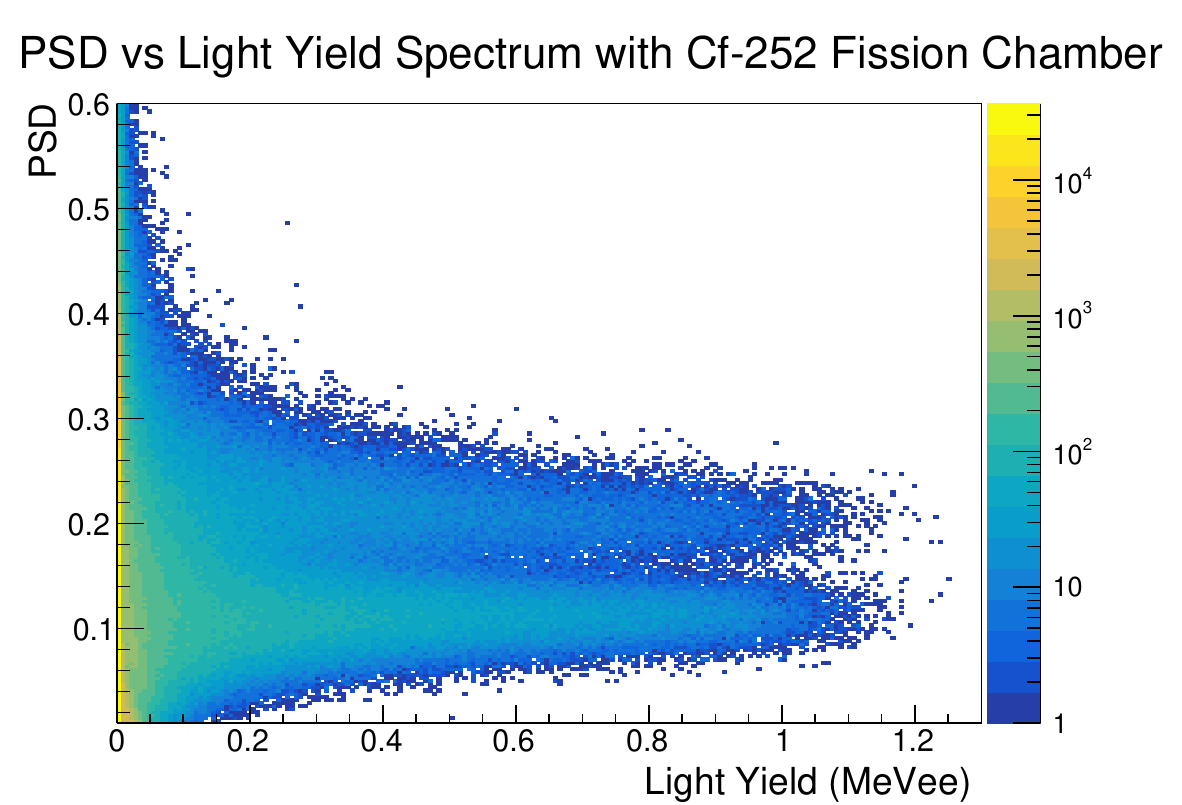}
    \caption{This plot shows the PSD for the PEN scintillation detector used in the fission chamber measurements. The neutron events and therefore proton recoils can be seen in the top band and the gamma events and therefore electronic recoils can be see in the bottom band.}
    \label{fig:FC_PSD_Plot}
\end{figure}

\begin{figure}
    \centering
    \includegraphics[width=0.7\textwidth]{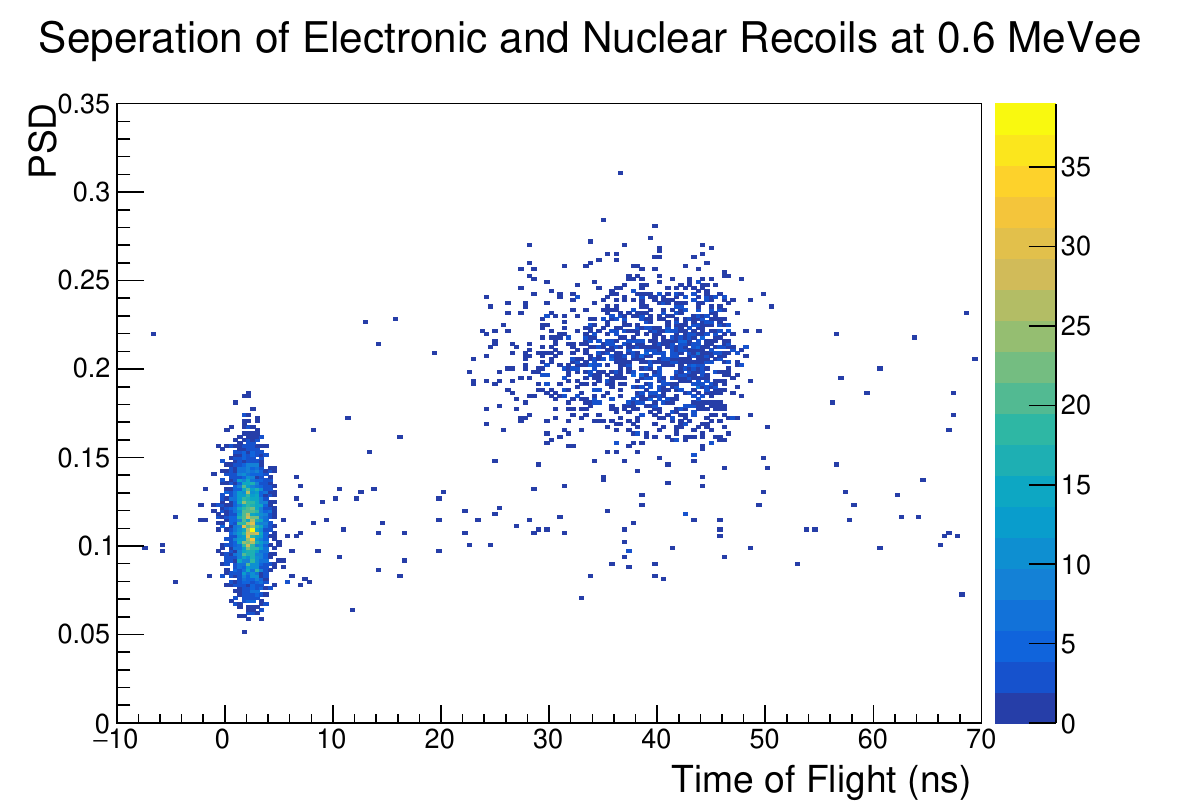}
    \caption{Separation of nuclear and electronic recoil events is demonstrated using a combination of time of flight and PSD for events with a light yield of $0.6 - 0.65$~MeV.}
    \label{tof_fc}
\end{figure}

The trigger from the ionization chamber is fed into a CAEN desktop digitizer model 5730, and coincidence is required between the fission chamber and PEN scintillation detector to record the event. 
The time of the trigger was defined by using the CAEN onboard Leading Edge Discriminator, or LED~\cite{caen_2020_White_paper}. 
Though using an LED may result in edge walking for varying pulse amplitudes, this was not a concern for the fission chamber pulses, as the rise time was faster than the digitizer sampling speed. 
The PEN scintillation detector had pulses with significantly slower rise times, therefore on board CFD (constant fraction discriminator), triggering at $50\%$ amplitude, was used to determine the start time of the pulse. 
The distance between the fission chamber and the PEN scintillation detector was measured to be $D = 0.890 \pm 0.005$ m. 
The gamma flash time of flight in PEN was centered at $\delta T = 2.8$~ns, with a resolution of $FWHM = 2.3$~ns (Fig.~\ref{fig:TOF_PLOTS}). 
This resolution was used to determine the neutron energy resolution, as shown in Eq.~\ref{kineticenergy}. 
This is very close to the resolution of channel-to-channel timing using the onboard CAEN DPP-PSD firmware ($\sim2.2$ ns)~\cite{caen_2020_White_paper}.
The total pulse was integrated over 430~ns, with the tail integral starting at an offset of 70~ns from the peak amplitude of the pulse~(Fig.~\ref{fig:FC_PSD_Plot}).
The identification of nuclear and electronic recoils is confirmed in Fig.~\ref{tof_fc}, where clear separation can be seen with PSD and time cuts for a light yield of $0.6 - 0.65$~MeV. 
The electronic recoil events can be seen to be clustered around 3~ns and the nuclear recoils are seen clustered significantly later at a higher PSD value. 

The light response of nuclear recoils in the detector were determined using a similar method as the calibrations, with the intention to have a consistency between the two data sets ($L(E)_{er}$ and $L(E)_{nr}$).
Energy deposited by neutrons in hydrogen-based scintillators has a step shape, that can be fit with a sigmoid function. 
The experimental data was therefore fit with a sigmoid function to determine the spectral shape (Eq.~\ref{sigmoid}). 
\begin{equation}
    \label{sigmoid}
    f(x) = \frac{p_0}{1+e^{-p_1(x-p_2)}+x*p_3}.
\end{equation}
The simulated data was then smeared using the resolution parameters defined in the calibration ($\alpha$ and $\beta$) and the smearing function in Eq.\ref{eq:sigsmear}: 
\begin{equation}
    \label{eq:sigsmear}
    E' = E + {erf}^{-1}(r)*\sqrt{2\left(\frac{E*\sqrt{\alpha^2+\frac{\beta^2}{E}}}{2.35}\right)^2}.
\end{equation}
An optimal quenching factor was then determined by scaling the smeared spectrum by a scalar, QF, and recreating the spectral shape by the sigmoid fit of the experimental data. 
In these fits, the parameters $p_1$ and $p_2$ were fixed and and $p_0$ and $p_3$ allowed to float. 
The accuracy of the quenching factor was defined by the variation of the QF as the $\chi^2$ per degrees of freedom, or $\Delta \frac{\chi^2}{DOF} = 2$.

The range of neutron energies measured from the \cf source was $0.5\leq E_n \leq2.0$ MeV. 
The width of the energy step for the fission chamber measurements was determined by the timing resolution of the experiment, or the FWHM of the gamma pulse.
This is equivalent to a FWHM resolution of 36 keV for a 1 MeV neutron. 
The energy steps for neutrons had a range that was larger or equivalent to the energy resolution from the timing.

\subsubsection{DD Generator}

A DD generator uses the fusion of deuterium to produce neutrons, as described in Eq.~\ref{ddequation}: 
\begin{equation}
    \label{ddequation}
    ^2H+^2H \to n + ^3He.
\end{equation}
This reaction releases monoenergetic neutrons at $E_n=2.45$ MeV~\cite{DD_generator_spectrum}. 
Because this source is monoenergetic, time-of-flight was not needed to determine the energy of the neutrons. 
The data was collected on the same CAEN 5730 desktop digitizer used in the \cf fission source measurements. 
The total pulse was integrated over 430~ns, with the tail integral starting at an offset of 70~ns from the peak amplitude of the pulse.

\begin{figure}
    \centering
    \includegraphics[width=0.7\linewidth]{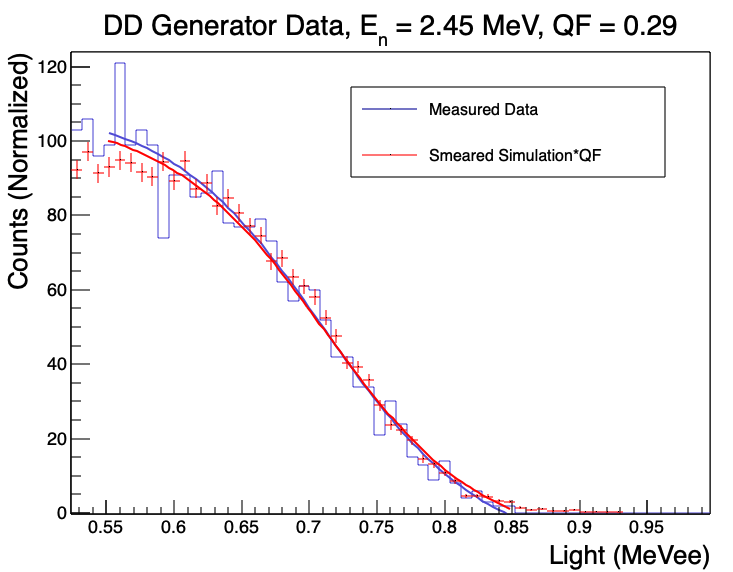}
    \caption{The measurement of neutrons from a DD generator is shown in blue, and fit with a sigmoid function, also in blue (Eq.~\ref{sigmoid}). The smeared simulation for neutrons of $E_n =2.45$ MeV is plotted in red, and scaled to the statistics of the data. The red line shows the simulation fit to the same sigmoid function as the data. }
    \label{fig:DD_plot}
\end{figure}

The data for the DD measurement was also fit with the sigmoid function (Eq. \ref{sigmoid}). 
The simulation data was generated in Geant4, and the energy deposited on protons from 2.45 MeV neutrons was recorded. 
The simulation was then smeared as described in Eq.~\ref{eq:sigsmear} and the spectral comparison is shown in Fig.~\ref{fig:DD_plot}. 

\subsubsection{Nuclear Reaction Neutrons}

\begin{figure}
    \centering
    \includegraphics[width=0.7\linewidth]{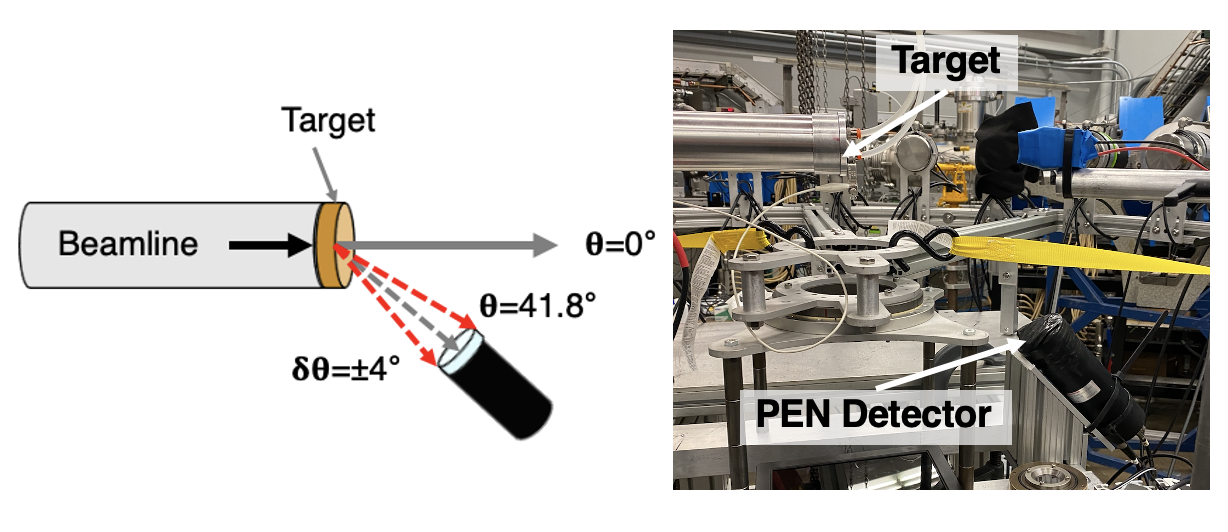}
    \caption{Experimental setup at University of Notre Dame Nuclear Science Laboratory with a diagram (left) and image (right). Detector was placed 30 cm from the target, at an angle of 41.8$^\circ$ from the beamline. }
    \label{fig:exp_setup}
\end{figure}

The nuclear reactions used in these measurements were \Lipn and \Can, and were conducted at the University of Notre Dame Nuclear Science Laboratory. 
The Sta. ANA 5 MV accelerator was used to produce a beam of $^1$H$^+$, $^4$He$^{+}$ or $^4$He$^{++}$, which was impinged onto a water-cooled target. 
For the \Lipn measurement, a beam of $^1$H$^+$ was impinged on a thin LiF foil (natural abundance) mounted onto a thick tantalum backing. 
For the \Can reaction, a beam of either $^4$He$^{+}$ or $^4$He$^{++}$ was impinged on a 99\% isotopically enriched $^{13}$C layer evaporated on a thick tantalum backing. 
The outgoing neutrons for both reactions were measured at angles $\theta = 41.8^{\circ}$ or $127.5^{\circ}$ in the laboratory frame. 
The experimental set up at $\theta = 41.8^{\circ}$ is shown in Fig.~ \ref{fig:exp_setup}.

\begin{figure}
    \centering
    \includegraphics[width=0.7\linewidth]{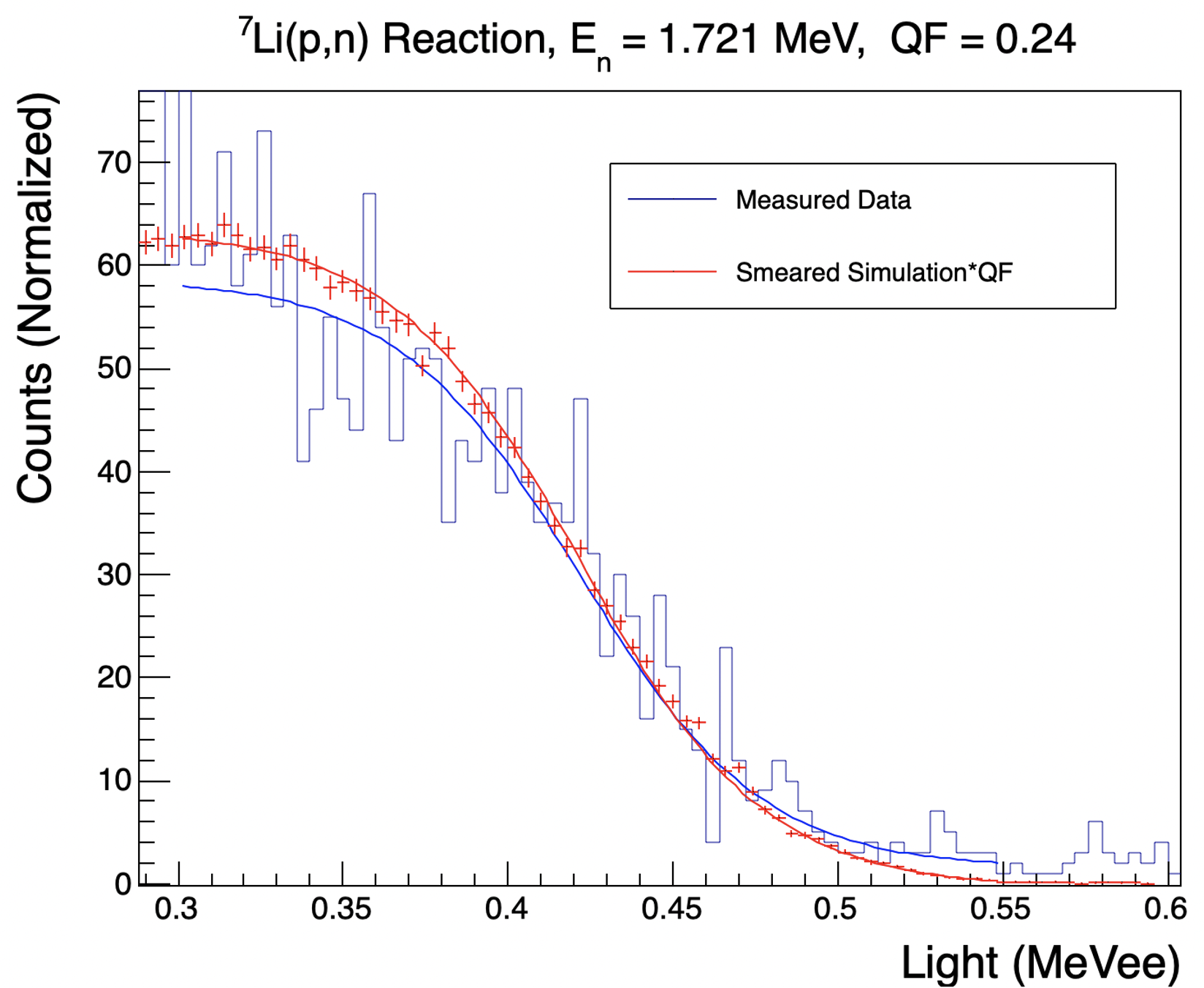}
    \caption{The data from $E_n=1.724$ MeV is plotted in blue with a sigmoid function (Eq.~\ref{sigmoid}). The simulation is plotted in red, fit with a sigmoid function and normalized to the statistics of the data. The data has excellent agreement with the simulation over the range where the sigmoid function is fit.}
    \label{fig:ND_neutrons}
\end{figure}

\begin{figure}
    \centering
    \includegraphics[width=0.7\linewidth]{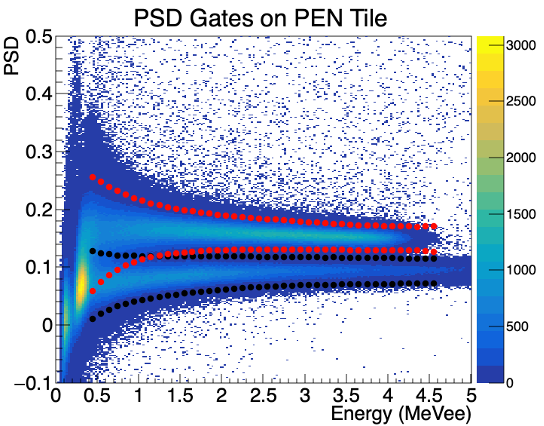}
    \caption{This plot shows the PSD gates for the PEN scintillation detector. Gates were used to distinguish neutron and gamma-ray events with an estimated probability of $2\sigma$. }
    \label{fig:ND_PSD}
\end{figure}

Waveforms taken from the PEN scintillation detector were recorded using a CAEN V1725 250 MS/s, 14-bit waveform digitizer. 
The total pulse was integrated over 440~ns, with the tail integral starting at an offset of 80~ns from the peak amplitude of the pulse.

s
The \Lipn reaction was used to measure neutrons in the range of $0.8\leq  E_n \leq1.9$~MeV and the \Can reaction was used to measure neutrons in the range of $2.5\leq E_n \leq7.1$ MeV.
The energy deposited by an equivalent energy neutron was simulated using Geant4 for each measurement.
As described in Section \ref{cf_section}, the data was fit with a sigmoid function, the terms $p_1$ and $p_2$ were fixed, and the sigmoid function was fit to the smeared neutron simulation. 
In Fig.~\ref{fig:ND_neutrons}, the data from $E_n = 1.721$ MeV is plotted with the corresponding simulation. 
The PSD gates were determined by projecting a light bin onto the PSD axis and fitting a double Gaussian for each the proton and electron recoil band (Fig.~\ref{fig:ND_PSD}). 
The PSD gates were then defined to be $2\sigma$ from the mean of the Gaussian for each recoil band.
The use of the PSD gates were vital in minimizing background events in the nuclear reaction measurements as the gamma-ray background was often significant.
An example is in the \Can reaction, if the beam energy is $E > 3.9 $~MeV, then the second excited state of $^{16}$O is populated, and will result in the emission of a $\sim6$ MeV gamma-ray. 
This high-energy gamma-ray is one example of potential backgrounds that could effect the measurements.

\section{Error Analysis}
\label{analysisandresults}

The accuracy of the neutron energy determination for the time of flight measurements with the fission chamber were determined by measuring the spread of energies for a given time and therefore energy slice. 
This was done by calculating the root mean squared of all the neutron energies of a certain slice:

\begin{equation}
    \label{root_mean_square}
    RMS = \sqrt{\frac{\sum_{i=1}^{n} \left( E_i-E_{Avg}\right)^2}{n} } .
\end{equation}

The accuracy in the energy for the nuclear reaction measurements was partially determined by the angular spread and therefore the kinematic acceptance of the detector geometry. 
Angular biases were deemed negligible as the detector was aligned with a well characterized setup frequently used at the facility~\cite{PhysRevLett.125.062501}.
It was also impacted by the variation seen in the beamline energy impinging on the target, which was less than 40 keV between runs. 

The error in the quenching factor values have been determined by variation seen in $\Delta \frac{\chi^2}{DOF}=2$, where $\chi^2$ is found by fitting the sigmoid function to the neutron edge. 

\section{Results}
\label{conclusion}

\begin{figure}
    \centering
    \includegraphics[width=0.7\linewidth]{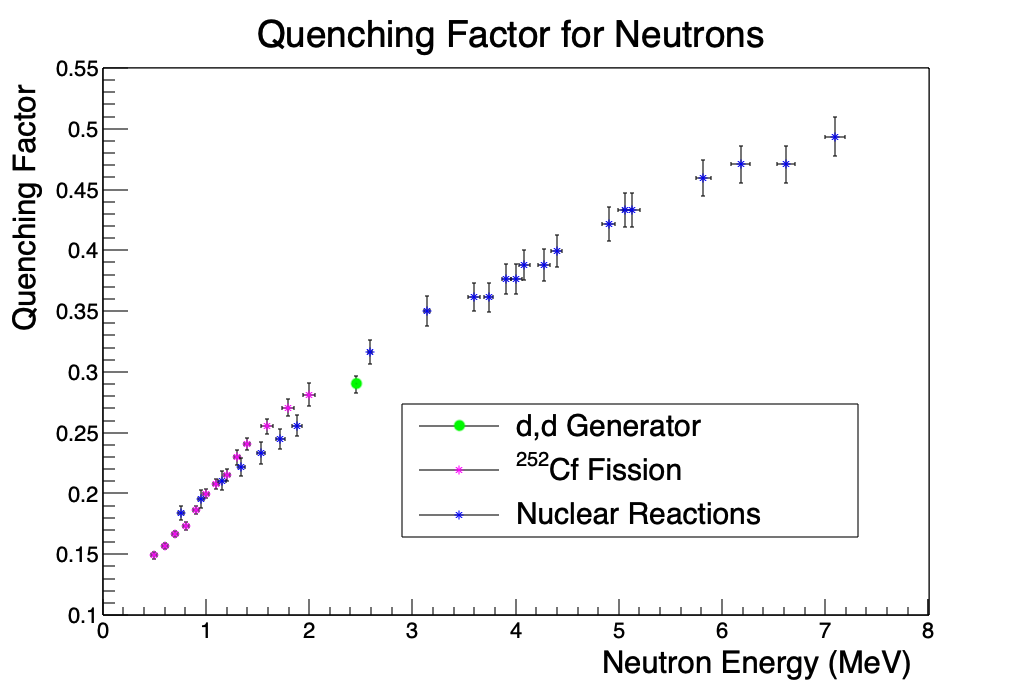}
    \caption{This plot shows the quenching factor of PEN as a function of energy. Quenching factors of all data show good agreement with each other. }
    \label{fig:QF_all}
\end{figure}

The quenching factors are plotted for the three measurement methods in Fig.\ref{fig:QF_all}. 
It is seen that all three measurements have a good agreement with each other and follow a similar trend.
At $E_n = 1.2 \pm 0.03 $~MeV, PEN has a quenching factor $QF = 0.215 \pm 0.007$. 
Compared to the commercial scintillator EJ-228, at $E_n = 1.21 \pm 0.06$~MeV, its quenching factor is $QF = 0.177 \pm 0.008$~\cite{WELDON2020163192}. 
Investigation into potential mechanisms for reduced light quenching in PEN is beyond the scope of this paper, but it is notable that this is a beneficial property for PEN as a self-vetoing material. 
It suggests that for nuclear recoils, the percentage of light quenched in PEN will be less than it is in the commercial scintillator EJ-228.

By using numerical integration, it is possible to analytically fit Birks' law to quenching factor data using Eq. \ref{birks}.
A best fit for the PEN light response was found and values for the scintillation efficiency, S and the Birks' factor, kB were derived. 
The ROOT Minuit fitting function determined the parameters which provided a minimum $\chi^2$ and their respective error, for a global fit. 

\begin{figure}
    \centering
    \includegraphics[width=0.7\linewidth]{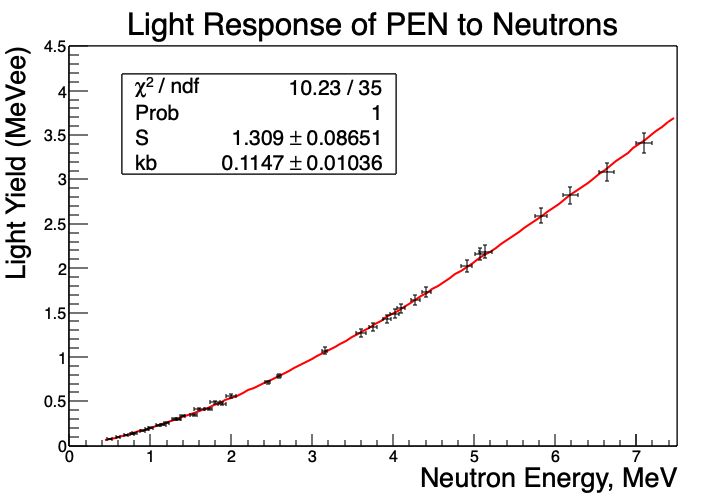}
    \caption{The light response of PEN, globally fit with Birks' function, as described in Eq. \ref{birks}. The paramters were found to be $kB = 0.12 \pm 0.01$~mm MeV$^{-1}$ and $S = 1.31\pm0.09$~MeV$_{ee}$~MeV$^{-1}$. }
    \label{fig:PENLY}
\end{figure}

By multiplying the quenching factors in Fig.~\ref{fig:QF_all} with the resepctive neutron energy (MeV), it is possible to generate the light response curve plotted in Fig.~\ref{fig:PENLY}. 
The light response curve is then fit with Birks' equation to generate the final parameters of $kB = 0.12 \pm 0.01$~mm MeV$^{-1}$ and $S = 1.31\pm0.09$~MeV$_{ee}$~MeV$^{-1}$. 
These parameters describe the shape of the quenching factor and its value for a given energy. 
The larger the $kB$, the more the light response becomes non-linear with higher proton recoil energy.

\begin{figure}
    \centering
    \includegraphics[width=0.7\linewidth]{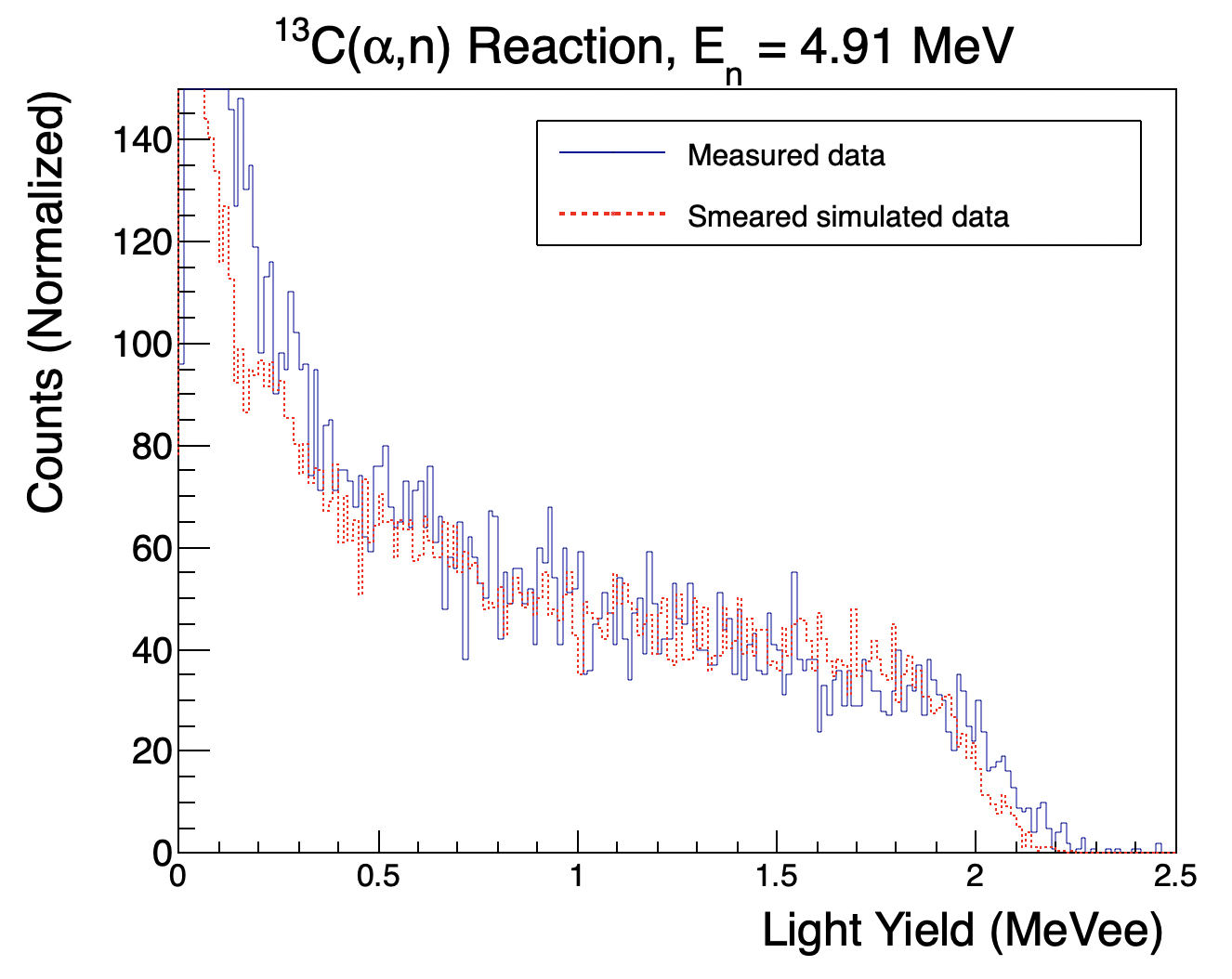}
    \caption{
    This figure shows data taken from neutrons generated in nuclear reactions, E$_n=4.91$ MeV. The simulated energy deposited by neutrons at 4.91 MeV are smeared and scaled given the light response of PEN, recreating the measured data.}
    \label{fig:dd_recreate_plot}
\end{figure}

The efficacy of the light response curve, and this comparison method, are demonstrated by comparing a smeared simulated spectrum with the response of PEN to neutrons at $E_n = 4.91$~MeV.
The experimental data was taken from the \Can reaction and the simulated spectrum was generated using Geant4. 
By defining the Birks' factor and scintillation efficiency, it is possible to define quenching factors for PEN at all energies.
In Fig.~\ref{fig:dd_recreate_plot}, each simulated energy deposited by the neutrons are smeared and scaled given the light response of PEN. 
There is strong agreement of the two spectra over the whole energy range, and the method used in this analysis does well to recreate the measured data. 

\section{Conclusion}

PEN has been identified as a material of interest for ton-scale physics experiments, for its inherent scintillation at $\lambda \approx 450$~nm and its wavelength shifting capabilities for VUV light. 
For applications in quasi background-free environments, PEN must act as a self-vetoing material and the light response of PEN is  necessary information to understand a PEN component's veto efficiency and therefore its effectiveness. 
The use of multiple neutron sources allows for the measurement of the light response of PEN over a large neutron energy range, with each measurement method having its own benefits. 
Pulse shape discrimination was used to separate electronic and nuclear recoils in the detector. 
This discrimination was most impactful when using the accelerator as a source of neutrons, due to the large number of gamma-ray events related to the reaction. 
The Birks' law was fit to the light response of PEN, Birks' factor was determined to be $kB = 0.12 \pm 0.01$~mm MeV$^{-1}$ and the scintillation efficiency is $S = 1.31\pm0.09$~MeV$_{ee}$~MeV$^{-1}$. 
With these parameters determined, it is now possible to evaluate PEN for low-background and ton-sale applications and to determine its viability in vetoing background events.

\section*{Acknowledgements}
This research was partially supported by the National Science Foundation through Grant No. PHY-1713857 and PHY-2011890, and the Joint Institute for Nuclear Astrophysics through Grant No. PHY-1430152 (JINA Center for the Evolution of the Elements).
It was partially supported by NKFIH grant No. K134197 and provided by the U.S. Department of Energy, Office of Science, Office of Nuclear Physics, under Award Number DE-AC05-00OR22725 and DOE DE-SC0014445.
This manuscript has been authored in part by UT-Battelle, LLC, under contract DE-AC05-00OR22725 with the US Department of Energy (DOE). The publisher acknowledges the US government license to provide public access under the DOE Public Access Plan (http://energy.gov/downloads/doe-public-access-plan).



\bibliographystyle{elsarticle-num}

\bibliography{main}

\end{document}